# Towards the Next Generation of Data Warehouse Personalization System
# A Survey and a Comparative Study

Saida Aissi[1], Mohamed Salah Gouider[2]

Bestmod Laboratory. University of Tunis, High Institute of Management of Tunis, Tunisia

[1]*saida.aissi@yahoo.fr*, [2] *ms.gouider@yahoo.fr*

**Abstract**

Multidimensional databases are a great asset for decision making. Their users express complex OLAP (On-Line Analytical Processing) queries, often returning huge volumes of facts, sometimes providing little or no information. Furthermore, due to the huge volume of historical data stored in DWs, the OLAP applications may return a big amount of irrelevant information that could make the data exploration process not efficient and tardy. OLAP personalization systems play a major role in reducing the effort of decision-makers to find the most interesting information. Several works dealing with OLAP personalization were presented in the last few years. This paper aims to provide a comprehensive review of literature on OLAP personalization approaches. A benchmarking study of OLAP personalization methods is proposed. Several evaluation criteria are used to identify the existence of trends as well as potential needs for further investigations.

## 1. Introduction

A data warehouse (DW) is defined as a collection of subject-oriented, integrated, non-volatile and time-variant data supporting management's decisions-making process [1]. The conventional DWs are designed based on a multidimensional view of data. It is based on a fact table representing the subject orientation and the focus of analysis.

The fact table contains usually numeric data called measures representing analysis needs in a quantified form. These measures are explored from different perspectives through dimensions.

OLAP systems allow dynamic manipulations by decision-maker of data contained in a DW. In fact, big volumes of historical data in DWs are analyzed using operations such as Roll-up and Drill-down. Drill-down operations allow the user to start from a general view of data in order to obtain a detailed view. Inversely, the Roll-up operation allows transforming detailed measures into summarized data.

DW store generally important quantity of information and multidimensional structures become increasingly complex to be understood at a glance. Even if a data mart is used, these structures still too complex. That is way, getting the required information become cost and tard and decision makers using OLAP tools may get frustrated.

Personalization aims to provide quick access to relevant information by eliminating all irrelevant information tailored to the needs, behaviors and user preferences. The goal of personalization is to deliver to most relevant information to the users in the most appropriate format and layout.

In this paper, we provide a literature review of developed and suggested proposals in the domain of OLAP personalization approaches and we critically compare and evaluate them in terms of several criteria, in order to identify the trends as well as the needs for further research in the area.

The remainder of this paper is organized as follows: Section 2 introduces the concepts of personalization and recommendation in DW systems. Section 3 presents an overview of several different approaches presented in the field of DW personalization as well as new concepts that have emerged in this domain. Section 4 presents a comparative study that provides a general, comparative view of the different approaches that have been presented. Section 5 presents a discussion and section 6 concludes the paper.

## 2. Adaptation and Recommendation Concepts

In the domain of OLAP personalization, we distinguish two main research orientations: (i) DW adaptation approaches and (i) DW recommendation approaches. These two concepts are introduced in this section.

*Adaptation*: We define adaptation as the process of adapting the system to the user needs, preferences, characteristics and requirements. System adaptation is usually related to defining and exploiting a user profile used to configure or adapt the system. Adaptation process aims to provide to the user the most relevant information in the most appropriate format and layout.

According to the type of adaptation action, we classify adaptation approaches as follows:

- Schema adaptation: This type of adaptation concerns the conceptual level. The personalization system adapts the conceptual schema according to the user's needs. Schema adaptation is realized by adding new hierarchy level [2, 3, 4, 5] or by filtering facts and/or dimensions [6, 7, 8].
- Visual adaptation: This type of adaptation concerns data visualization. The system presents personalized visualization of the response to OLAP query according to the preferences of the user and taking into account visual constraints [9, 10].
- Adaptation of analyzes: This type of personalization aims to adapt the OLAP query expression according to the user needs and preferences. This is realized by proposing new preference algebra allowing the user to express his preferences over the MDX queries [11, 12, 13] or by adapting the content of the multidimensional tables according to the user needs [14, 15].

***Recommendation***: We define recommendation in OLAP systems as the process that proposes a new OLAP query to the user according to his preferences and needs in order to facilitate the analysis process and assist the user during the exploration of the OLAP system.

Giacometti et al [16, 17] and Jerbi et al [18, 19] define recommendation as a process that exploits user's previous queries on the cube and what they did during the previous session in order to recommend the next query to the actual user. In recommendation process, the recommended query is different from the initial query due to different user interests.

In the domain of OLAP recommendation, we distinguish two main research orientations:
- Collaborative recommendation approaches based on query log analyzes: Based on the current user query and the past navigation log, the system recommend alternatives queries to help users navigating the cube [16, 17, 20, 21]
- Individual recommendation approaches based on user profile analyzes: the system provides alternatives and anticipated recommended query taking into account the user context [18, 19].

## 3. Survey of OLAP Personalization Approaches.

This section presents a thorough survey on the proposed approaches in the domain of DW personalization. The existing approaches are classified according to the two mains researches strategies distinguished in the domain: *adaptation strategies* [2-15] and *recommendation strategies* [16-20].

### 3.1. DW adaptation approaches

We classify DW adaptation approaches according to aim of the adaptation action, into four main types: (i): *schema adaptation approaches* [2- 8],*(ii) visual adaptation approaches* [9, 10]*, and (iii) analyzes adaptation approaches* [11- 15].

#### 3.1.1 Schema adaptation approaches

Garrigos et al [6] propose to customize the schema of the DW at the conceptual level. The personalized OLAP schema is adapted to the specific characteristics of the decision maker. The approach is based on a:
1. User model that contains all user related information as his characteristics, context, requirements and behaviors.
2. A set of personalization rules specified using the PRML language (Personalization Rules Modeling Language). The personalization rules allow specifying the required personalization actions.

The personalization process is based on the following ECA principle: On Event, If Condition Then Action. The tracking events are possible OLAP operations that manipulate multidimensional structures (eg; *AddDimension, RemoveDimension, Rollup, DrillDown,…*).

The defined personalization actions consists in filtering and selecting facts, dimensions, attributes and aggregation functions according to the user model and the set of PRML rules.

The last presented approach is extended by Glorio et al [7, 8] in order to integrate spatiality in the personalized OLAP schema. When building the multidimensional model, the designer defines some spatial personalization rules using a PRML language (Personalization Rules Modeling Language).

The multidimensional model is personalized according to the defined personalization rules. The spatial personalization actions consist in changing the structure of the DW, adding geometric description to the multidimensional elements (*BecomeSpatial* action) or adding geometric elements (*addLayer* action). Actions of personalization introduce spatiality in the multidimensional model and allow by consequence the execution of spatial analysis over the DW.

Favre et al [2, 3, 4] propose a collaborative DW adaptation approach for the enrichment of analysis in XML data warehouses. The adaptation process consists in personalizing the DW schema by creating a new hierarchy level or by enriching an existing one. The personalization actions are performed according to the user knowledge explicitly expressed through if-then rules. The aggregation rules are composed of a fixed part and an evolving part. The fixed part presents global analysis needs and is relied to the warehouse schema. The evolving part is composed by the aggregation rule creating new granularity levels on the dimension hierarchies. The implementation of the approach generated the WEDriK plateform [2].

This approach has the advantage of sharing the new scheme of DW between users in a collaborative environment. However, the approach depends on explicit extraction of the user knowledge which may disturb the user and slow the personalization process. The proposals of Garrigos and Glorio et al [6, 7, 8] are richer than the proposal of Favre et al [2, 3, 4] relative to the personalization factors. In fact, In [6, 7, 8] the approach proposes an adaptation of the multidimensional schema taking into account the user characteristics, context, needs and behavior, However, the approach of [2, 3, 4], presents an adaptation of the schema basing only on the user's knowledge.

Bentayeb et al [5] extend the proposal of Favre et al [2, 3, 4, 5] propose to create the new hierarchy level after recommending and validating the appropriate adaptation by the user. Recommendation action is based on proposing a new OLAP operator (RoK) using Data Mining method (the k-means method). In fact, the user define the algorithm parameters, the system will extract and recommend the appropriate cluster to the user, Once the cluster is validated by the user, it will be integrated in the DW by creating a new hierarchy level.

The approaches of Favre et al [2, 3, 4] and Bentayeb et al [5] depends essentially on explicit user implication in different stages: In the knowledge acquisition phase, in defining the algorithm parameter and in validating the recommended cluster. Thus could slows the adaptation process and disturb the user from his objectives.

The approaches could be classified as individual approaches as the schema evolution is realized according to the needs of each user and it could be classified as collaborative approaches in the sense that the adapted schema could be shared by many users. The DW is updated, allowing to share the new analysis possibilities with all decision makers [5]. A problem arises when the user no longer need those new levels.

Khemiri et al [22] propose an approach based on the idea of creating a materialized view of DW according to each user profile. The user profile is composed by a set of permanent preferences (called rigid preferences) and temporary preferences (called flexible preferences). The personalized materialized view is adapted to rigid preferences. Instead of consulting all DW, when the user launches a query, the system uses the materialized view and integrates the flexible preferences to answer. However, the process of creating and adapting a materialized view to each user is cost and complicated.

### 3.1.2 Adaptation of analyzes

A second research direction in the field of OLAP adaptation deals with the customization of analyzes.

Golfarelli et al [11] and Biondi et al [12] propose to adapt the expression of MDX queries according to the user preferences. The approach allows the expression of preferences over the OLAP query. An OLAP algebra (MYolap algebra), which allows users to declare their preferences on, attributes, measures and aggregation level is proposed. For Evaluating preferences, an algorithm (WEST algorithm) which handles preferences on categorical and numerical data as well as on aggregation level of data is proposed.

The proposed approach is an *individual adaptation approach based on Preference Construction*. Indeed, preferences are formulated on attributes, measures and hierarchies through a defined algebra.

Aligon et al [13] extend this approach by applying association rules on the log of MDX queries to extract implicitly user preferences and to annotate the queries of the user with the MYOLAP algebra. In this approach the user preferences are extracted without user intervention using mining technique. The adaptation process is based on three main steps: First, using data-minig techniques, a set of association rules that relate sets of frequent query fragments is extracted for the log of past MDX queries of the current user. Second, when the user launches a query, a subset of pertinent and effective rules is selected. Finally, the user's query is annotated by preference resulting from the translation of the set of selected rules.

In order to customize the analyses, a personalization approach based on preferences construction is proposed by Ravat et al [14] [15]. The content of the multidimensional tables are adapted basing on a set of user preferences which are expressed as a priority order on the attributes of the multidimensional model.

In this context, two approaches are proposed: the naive approach which's based on according a weight to the different attributes of the model in order to apply a priority order during their visualization. And the advanced approach, where the priority order of the different attributes varies according to the usage context and manipulation operations using an ECA configuration. Only attributes which weight exceeds a threshold are displayed.

However, fixing a minimum threshold to present attributes may reduce the number of eventual analysis operations that the user may need. The eliminated attributes should be mentioned in order to be presented if the user needs them.

Finally, Thalhammer [23] propose architecture for active DW in order to automate decision making tasks. The authors propose to exploit the results obtained during analysis in order to improve the data preprocessing and increase performance. The proposal is based on the definition and the implementation of a set of ECA mechanisms.

### 3.1.2 Visual adaptation approaches

Another research orientation in domain of individual OLAP adaptation concerns the adaptation of the response to OLAP query or visual adaptation.

Bellatreche et al [9, 10] propose a framework for visual personalization of the response to MDX queries taking into account not only visualization constraint but also user profile. The aim of the approach is to provide the best relevant visualizations of the query answer to the user. User preferences are presented through a preference model based on partial order expressed explicitly by the user on dimensions and members of the cube. The response to any query may contain measures with different levels of aggregation. In addition, several members of dimensions can be combined.

## 3.2. DW recommendation approaches

In the domain of DW recommendation approach, we distinguish (i) collaborative recommendation approaches exploiting user session analysis [18-21] and (iii) individual recommendation approaches exploiting user profile analysis [16, 17].

### 3.2.1. Collaborative recommendation approaches exploiting user session analysis

An approach for query recommendation is proposed by Sapia [20, 21]. The recommendation method is based on a probabilistic Markov model. The approach could be summarized in the following steps: First, a prototype of query is built for each query in the log sessions. Second, prototypes of queries are compared basing on a distance that gives the number of operations to move from one prototype to another. Then, a Markov model predicts the probability of occurrence of each prototype. And finally, the prototypes of queries with the highest probability of occurrence are used for the recommended query.

The approach takes into account the sequence of queries and the fact that the database is exploited by many users.

Giacometti et al [18, 19] propose a collaborative approach to recommend OLAP queries expressed in MDX language. A measurement of the distance of two MDX queries as well as a measurement of the distance of two sequences of MDX queries is proposed. Once the user launches a query, the system, searches the log of the OLAP server in order to find the candidate sessions that resemble the most to the actual session. Once candidate sessions are selected, the distance between the last query of the current session and the last query of each candidate session are calculated. Finally, the query that resembles the most to the current query of the user will be proposed as a recommended query.

The last two presented approaches provide a recommendation system that takes into account user preferences implicitly learned through the user's interactions with the system.

The approaches are independent of the intervention of the user to specify his preferences, this makes the approach more efficient and objective.

However, these approaches do not take into account the context of use and user characteristics. Such criteria play a fundamental role in the success of recommender systems.

### 3.2.2. Individual recommendation approaches exploiting user profile analysis

Jerbi et al [16, 17] propose an individual *approach for OLAP query recommendation* based on the exploitation of the user profile. Indeed, they propose, first, to represent as a tree the current query of the user and the user profile. Second, a tree matching algorithm is used to compare the two trees (Tree of the user query and the tree of the user profile). The third step consists in adding to the tree of the query, the parts of the tree of preferences (nodes and arcs, which do not appear). Finally, the query resulting from the process of transformation of the tree of the initial query is proposed as a recommendation to the user.

The method proposed by [16, 17] can guide and make recommendations to enhance the current request of the user according to their preferences (from customizing profile). However, this method does not take into account previous queries already posed by the user.

## 4. Comparing OLAP personalization approaches

The following section presents a comparative study that provide a general, comparative view of the different approaches that have been presented and discussed in the field of DW personalization. The different models are compared against these criteria.

*Personalization factors*: The first criterion of comparison included in our comparative study is the factors of personalization. In fact, according to [6] personalization is based on several factors like user characteristics, user context, user behavior and user requirements. We add to those factors, the user preferences.

   a) **User characteristics**: This factor presents the specific characteristics of the user, we consider the age, the user role, the user knowledge and the language.
   b) **User context**: It represents the information related to the spatial and temporal environment of the user when manipulating the system. User context includes location, time and the specific characteristics of the user device
   c) **User behaviour:** The user behavior represents information related to the user manipulation of the system. (e.g; OLAP manipulations, spatial selection…). The user behavior is an implicit way to extract the user preferences.

d) **User preferences:** This criterion represents the user preferences extracted explicitly through direct questioning of the user. The preferences could be expressed by different manners (e.g; Through a priority order granted to members and dimensions of the schema [9, 10], through a weight to schema structure [14, 15], through a preference algebra [11, 12, 13])

e) **User requirements:** this criterion is related to some specific requirements or expectations of the Decision makers when manipulating the system. We consider requirements in terms of security, configurations and performance tuning.

*Approach orientation*: this criterion indicates if the proposal presents a DW personalization approach or a DW recommendation approach.

*Nature of the approach*: the third criterion included in the comparison is the nature of the approach proposed by researchers. Indeed, the approach of personalization could be based and centered on one user, we speak then about an individual approach as it can be based on the behavior of a group of users, and we speak then about collaborative approach.

*User profile:* this criterion indicates if the personalization approach is based, or not, on the definition of the user profile. It indicates also if the profile is detected implicitly through the user manipulations of the system or explicitly by taking directly information from the user.

*Type of adaptation action*: the other criterion of comparison included in our study is the type of adaptation action. We define three types of adaptation action: (i) schema adaptation, (ii) adaptation of the answer to the user query (or visual adaptation) and (iii) adaptation of the expression of OLAP query (e.g; by proposing an algebra allowing the expression of the user preferences).

*Type of recommendation action:* In the context of a recommendation approach, criteria specify the nature of the recommendation action (e.g; recommendation of an MDX query, recommendation of a cluster…).

Table 1 reports a comparison of the above approaches according to the presented criteria.

## 5. Discussion and perspectives

We distinguish two major line of research in the domain of OLAP personalization: (i) researches proposing specific recommendations to the user in order to facilitate and accelerate the analysis process [16- 20] and (ii) researches offering an adapted OLAP system customized to the specific characteristics of the user profile (preferences, behavior, requirements...) [2-14, 23].

We identified three main research orientation for introducing adaptation in OLAP systems: (i) schema adaptation approaches [2-8], (ii) adaptation of analyzes [8, 9] and (iii) visual adaptation approaches [9, 10].

Regarding OLAP recommendation systems, we have highlighted two major research strategies: (i) collaborative recommendation strategies exploiting user session analysis [18-21] and (ii) individual recommendation strategies exploiting user profile analysis [16, 17].

We have identified user preferences as the most important factor in building recommendation and adaptation systems. Indeed, all proposed personalization approaches reflect this criterion in developing the customization process.

A first possible research field concerns the construction of the user profiles in DWs. In fact, in OLAP personalization process, the user preferences are extracted either explicitly through user intervention [2-12, 14-17, 22] or implicitly through user interactions with the system. [13, 18, 19, 20, 21, 19]. Most approaches are based on explicit extraction methods ( eg; weights on schema structure, order of priority on cube components...) However, using explicit profiling have some disadvantage, In fact, getting such information explicitly needs almost the interruption of the user work flows, as his needs and preferences should be stated explicitly by filling out a form or answering several questions that could disturb the user and slows its task. As a result, many users may skip or refuse this step. In such a case, no information can be build about the user's preferences and needs. Implicit profiling appears as an alternative solution for such a problem.

A possible research filed consists in taking advantage of learning techniques in order to learn the user behavior and predict his next system manipulation. Moreover, using classification and clustering techniques in order to detect similar user's profiles and behaviors could improve and accelerate the recommendation and personalization process and develop implicit profiling techniques.

Otherwise, making the collection of user profile components implicitly and transparently enables to have a more efficient customization process. Kozmina et al [24] propose a method for user-describing profile construction defining temporal, spatial, preferential, and interactional and recommendation user profile. Kobs et al [25] presented a state of the art on modeling of users based on system needs.

A second research field concerns the development of the customization factors. The user requirements are a personalization factor not well exploited in OLAP systems (the only work that considers this criterion is that of Bellatreche et al [8, 9]). The proposal of a personalization system that takes into account user requirements in terms of level of security, performance and configuration, is a line of research that has to be more exploited.

| | Proposal | | Ravat and al, 2008 | Bellatreche and al, 2005 | Favre and al, 2007 | Bentayeb and al,2009 | Giacometti and al,2008 | Glorio and al., 2010. Garrigos and al., 2009 | Jerbi and al., 2009 | Golfarelli, 2009. Biondi,2011 | Bellatrecheand, 2005,2006 | Thalhammer and al, 2001 | Aligon ,2011 | Sapia, ,20001 |
|---|---|---|---|---|---|---|---|---|---|---|---|---|---|---|
| **Personalization factors** | *Context* | Location | - | - | - | - | - | × | - | - | - | - | - | - |
| | | Time | - | - | - | - | - | × | - | - | - | - | - | - |
| | | device | - | × | - | - | - | × | - | - | × | - | - | - |
| | *Characteristics* | age | - | - | - | - | - | × | - | - | - | - | - | - |
| | | User role | - | - | - | - | - | × | - | - | - | - | - | - |
| | | language | - | - | - | - | - | × | - | - | - | - | - | - |
| | | khnowledge | - | - | × | × | - | - | - | - | - | - | - | - |
| | *Requirements* | Security | - | - | - | - | - | - | - | - | - | - | - | - |
| | | Performance tuning | - | - | - | - | - | - | - | - | - | - | - | - |
| | | User configuration | - | - | - | - | - | - | - | - | × | - | - | - |
| | *Behavior* | | - | - | - | - | × | × | × | - | - | - | - | × |
| | *Preferences* | | × | × | - | - | - | - | - | × | × | × | × | - |
| **Approach orientation** | *Adaptation* | | × | × | × | × | - | × | - | × | × | × | × | - |
| | *Recommendation* | | - | - | - | - | × | - | × | - | - | - | - | × |
| **Nature of the approach** | *individual* | | × | × | - | - | - | × | × | × | × | × | × | - |
| | *Collaboratif* | | - | - | - | - | × | × | - | - | - | - | - | × |
| **user profile** | *Implicit* | | - | - | - | - | × | - | - | - | - | - | × | × |
| | *Explicit* | | × | × | × | × | - | × | × | × | × | × | - | - |
| **Type of adaptation action** | *Schema* | | - | - | × | × | - | × | - | - | - | - | - | - |
| | *visualization* | | - | × | - | - | - | - | - | - | × | - | - | - |
| | *Analyzes* | | × | - | - | - | - | - | × | × | - | × | × | - |
| **Type of recommendation action** | *query* | | - | - | - | - | × | - | - | - | - | - | - | × |
| | *cluster* | | - | - | - | - | - | - | - | - | - | - | - | - |

**Table 1. Comparative study between surveyed approaches on OLAP personalization**

Jerbi et al [16, 17] propose a recommendation of annotated query in order to help the user in the system exploitation process. However, no works propose an adaptation of the system according to the user level of knowledge. This factor is completely omitted in OLAP personalization systems. Proposing a customization approach that allows an adaptation of the system to the user level of knowledge is a potential research field.

Moreover, neither approach includes all factors of customization in the personalization process user needs, preferences and requirements). User profile modification in the domain of SOLAP systems that could be more investigated in order to be enriched with more complete features of user profile as well as spatial interactions with the SOLAP system. The definition of a user profile that includes the full specifications covering all the requirements for presentation and interaction enables to make the customization process more complete and efficient.

Works dealing with personalization at the visualization level are presented essentially in the proposal of Belleatreche and al [8, 9]. This type of personalization aims to offer an adapted visualization of the response to OLAP queries taking into account visualization contraint as well as some user preferences on cube structure. Taking into account the various elements of the user profile as the spatial and temporal context and the different characteristics of the user (age, language ...) is completely omitted in this type of customization.

A third possible research field concerns personalization actions. No approach proposes a complete personalization process offering a personalization of the schema, interaction and visualization. The definition of a flexible approach allowing all types of personalization actions according to the user needs could be considered.

Moreover, it is estimated that 80% of data stored in databases has a spatial component [26], to enrich the analysis possibilities of geographic data, technologies such as spatial data mining, Spatial OLAP systems (SOLAP) and spatial DWs have been proposed. SOLAP systems and spatial DW provide new opportunities for multidimensional analysis of spatial information (Bedard et al., 2006). SOLAP users have specific needs, preferences and goals. However, SOLAP personalization is a search field not well exploited. In fact, the only work proposing personalization of SOLAP systems is proposed by Glorio et al. [7, 8] who present an adaptation of the DW schema by integrating the required spatiality at the conceptual level. However, Personalization of SOLAP systems at the visualization level is completely ignored.

SOLAP users rely almost on visual interaction with the SOLAP systems and cartographic presentation of the result of their queries. Very often, these responses can not be fully visualized and the user must navigate through them to find relevant data. Proposing personalized visualization of the cartographic data will make the analysis process more efficient and optimized. This personalization can affect the content of the map (visualizing only the interesting parts of the map according to the user needs) and the content of the multidimensional tables. For the same query, different users may obtain different visualizations according to their preferences.

Establishing recommendation process in the context of SOLAP systems is also a research filed not explored. SOLAP recommendation process could propose alternative and/or anticipatory recommended query and/or spatial maps and/ or data layers.

## 6. Conclusion

In this paper, we dress an overview of developed and suggested OLAP personalization approaches. Each approach is presented and discussed, then, a comparative study between the different proposed works is presented in order to compare and evaluate them in terms of some criteria. In this paper we have identified three main research strategies in the domain of OLAP adaptation: OLAP schema adaptation strategies, adapation of analyze strategies and visual adaptation strategies. In the domain of OLAP recommendation, we distinguished collaborative recommendation approaches exploiting user session analysis and individual recommendation approaches exploiting user profile analysis.

The proposed work allow us to have a global vision on different proposals and take advantages of the studied contributions in an optimized way in order to introduce our future work which is the proposal of a new approach on spatial DW personalization.